\documentclass[11pt]{article}
\usepackage{amsmath}
\usepackage{amssymb}
\usepackage{latexsym}
\usepackage{graphicx}
\usepackage{epsfig}
\usepackage{stmaryrd}
\usepackage{framed}

\def\Bbox{
{\unskip\nobreak\hfil\penalty50
\hskip1em\hbox{}\nobreak\hfil{\lower .5pt \hbox{$\Box$}}
\parfillskip=0pt \finalhyphendemerits=0 \par}
}

\def\eop{
\ifmmode {\hbox{\Bbox}} \else \Bbox \fi
}

\def\bbox{
\ifmmode {\hbox{\bbox}} \else \Bbox \fi
}

\newtheorem{thm}{\bf Theorem}
\newtheorem{lem}[thm]{\bf Lemma}        
\newtheorem{prop}[thm]{\bf Proposition}  
\newtheorem{cor}[thm]{\bf Corollary}

\newcommand{\id}{\mathbf{1}}

\begin{document}

\title{{\bf \Large Some Remarks on Conway and Iteration Theories}}

\author{
Zolt\'an \'Esik\thanks{Partially supported by 
grant no. ANN 110883 from the National Foundation of Hungary for
Scientific Research.}\\
Dept. of Computer Science\\
University of Szeged\\
Hungary
\and
Sergey Goncharov\\
Dept. of Computer Science\\
Friedrich-Alexander-Universit\"at Erlangen-N\"urnberg\\
Germany
}

\date{}

\maketitle

\begin{abstract}
We present an axiomatization of Conway theories which yields,
as a corollary, a very concise axiomatization of iteration theories 
satisfying the functorial implication for base morphisms. 
\end{abstract}

It has been shown that most fixed point operations in computer science 
share the same equational properties. These equational properties are captured by the 
notion of iteration theories \cite{BEW1,Esikaxioms}. Several axiomatizations of iteration 
theories have been presented in \cite{BEbook,Esikaxioms}. For a recent overview,
we refer to \cite{EsMFCS}. 

The axioms of iteration theories can be conveniently divided into two groups: 
axioms for Conway theories 
and the commutative identities. The commutative identities have later 
been simplified to the group identities or certain generalized 
power identities, cf. \cite{Esgroup,EsAC,EsMSCS}. In this note, we 
provide further axiomatizations of Conway and iteration theories,
see Corollaries~\ref{cor-Conway1},\ref{cor-Conway2} and 
\ref{cor-iteration}. 

Iteration theories having a  constructable fixed point operation 
(such as the theories of monotonic or continuous functions over a 
cpo or a complete lattice equipped with the least fixed 
point operation, or the theories of continuous functors over 
$\omega$-categories equipped with the initial fixed point operation) 
usually satisfy the `functorial implication' for base morphisms,
see \cite{BEbook}.
In fact, the commutative identities were introduced in \cite{Esikaxioms} 
as a strictly weaker, but fully equational substitute for 
the functorial implication for (surjective) base morphisms,
which are however still sufficient for completeness. As a corollary of our results,
we obtain a simple axiomatization of iteration theories with a functorial 
dagger for base morphisms, cf. Corollary~\ref{cor-weakiteration}.

We assume familiarity with Conway and iteration theories 
and closely follow the terminology and notation in \cite{BEbook}.

\begin{prop}
Let $T$ be a preiteration theory. Then $T$ is a Conway theory iff 
$T$ satisfies the (base) parameter, fixed point, permutation and double dagger 
identities as well as the following identity:
\begin{eqnarray}
\label{eq-specpairing}
(\id_n \oplus 0_m) \cdot \langle f\cdot (\id_n \oplus 0_m \oplus \id_p), g\rangle^\dagger 
&=& f^\dagger,
\end{eqnarray} 
where $f: n \to n + p$ and $g: m \to n + m + p$. 
\end{prop}

{\sl Proof.} 
It is known that all identities mentioned in the proposition hold in 
Conway theories, see the summary on p. 212 and Proposition 3.18 on p. 134 in \cite{BEbook}. 
Suppose now that the identities mentioned in the proposition hold. 
We prove that the pairing identity
\begin{eqnarray}
\label{eq-pairing} 
\langle f,g\rangle^\dagger 
&=& 
\langle f^\dagger \cdot \langle h^\dagger, \id_p\rangle,h^\dagger\rangle
\end{eqnarray} 
holds for all $f: n \to n + m +p$ and $g: m \to n + m + p$, where 
\begin{eqnarray*}
h &=& g \cdot \langle f^\dagger,\id_{m+p}\rangle: m \to m + p.
\end{eqnarray*} 
We follow closely the argument on p. 164 and 165 in \cite{BEbook}.
The difference is that we use (\ref{eq-specpairing}) instead 
of the `simplified composition identity'. 

First, we establish the pairing identity in two special cases. 
The first special case is
\begin{eqnarray*}
\langle f \cdot (\id_n \oplus 0_m \oplus \id_p), 
g \cdot (\id_n \oplus 0_m \oplus \id_p)\rangle ^\dagger
&=& 
\langle f^\dagger, g \cdot \langle f^\dagger, \id_p\rangle\rangle,
\end{eqnarray*}
where $f: n \to n + p$ and $g: m \to n + p$. Indeed, 
let $h = \langle f \cdot (\id_n \oplus 0_m \oplus \id_p), 
g \cdot (\id_n \oplus 0_m \oplus \id_p)\rangle$, then 
\begin{eqnarray*}
(\id_n \oplus 0_m) \cdot  h^\dagger
&=& f^\dagger
\end{eqnarray*} 
by (\ref{eq-specpairing}),
so that $h^\dagger = \langle f^\dagger, k\rangle$ for some 
$k: m \to p$. Hence, 
\begin{eqnarray*}
h^\dagger
&=& 
h \cdot \langle h^\dagger, \id_p\rangle \\
&=& 
\langle f \cdot (\id_n \oplus 0_m \oplus \id_p), 
g \cdot (\id_n \oplus 0_m \oplus \id_p)\rangle \cdot \langle f^\dagger, k, \id_p\rangle \\
&=& 
\langle f \cdot \langle f^\dagger, \id_p\rangle, g \cdot \langle f^\dagger, \id_p\rangle \rangle\\
&=& 
\langle f^\dagger, g \cdot \langle f^\dagger, \id_p\rangle \rangle
\end{eqnarray*}
by the fixed point identity.

The second special case is
 \begin{eqnarray*}
\langle 0_n \oplus f, 0_n \oplus g \rangle^\dagger 
&=& 
\langle f\cdot \langle g^\dagger, \id_{p}\rangle, g^\dagger\rangle,
\end{eqnarray*} 
where $f: n \to m + p$ and $g: m \to n + p$. 
This follows from the first special case using the 
permutation identity (or the block transposition identity which is an 
instance of the permutation identity). 

Now using the two special cases, the pairing identity (\ref{eq-pairing}) in full generality 
is established exactly as on p. 165 of \cite{BEbook}. 

Let $f: n \to n + m + p$ and $g : m \to n + m + p$, and define 
$h = g \cdot \langle f^\dagger,\id_{m+p}\rangle$ and $\tau = \langle \id_{n+m},\id_{n+m} \rangle$.
Then 
\begin{eqnarray*}
\langle f,g\rangle^\dagger 
&=& (\langle f \cdot (\id_n \oplus 0_{m+n} \oplus \id_{m+p}), 
             g \cdot (\id_n \oplus 0_{m+n} \oplus \id_{m+p})\rangle \cdot 
( \tau \oplus \id_p))^\dagger\\
&=& \langle f \cdot (\id_n \oplus 0_{m+n} \oplus \id_{m+p}), g \cdot (\id_n \oplus 0_{m+n} \oplus \id_{m+p})^{\dagger\dagger},
\end{eqnarray*}
by the double dagger identity, 
\begin{eqnarray*}
&=& \langle 0_n \oplus f^\dagger, g \cdot \langle 0_n \oplus f^\dagger, 0_n \oplus \id_{m+p}\rangle\rangle^\dagger,
\end{eqnarray*}
by the first special case and the (base) parameter identity, 
\begin{eqnarray*}
&=& \langle 0_n \oplus f^\dagger, 0_n \oplus h\rangle^\dagger \\
&=& \langle f^\dagger \cdot \langle h^\dagger, \id_p\rangle,h^\dagger\rangle,
\end{eqnarray*}
by the second special case. \eop

Suppose that $T$ is a preiteration theory and $\mathcal{C}$ is a set of morphisms in $T$. 
Following \cite{BEbook}, we say that $T$ satisfies the functorial implication 
for $\mathcal{C}$ if for all $f: n \to n + p$ and $g: m \to m + p$ in $T$
and $\rho: n \to m$ in $\mathcal{C}$, 
\begin{eqnarray*}
f \cdot (\rho \oplus \id_p) = \rho \cdot g &\Rightarrow & f^\dagger = \rho \cdot g^\dagger. 
\end{eqnarray*}

\begin{lem}
\label{lem-specpairing}
Let $T$ be a preiteration theory. Then the functorial implication holds in $T$ for 
injective base morphisms iff the permutation identity and (\ref{eq-specpairing}) hold. 
\end{lem}

{\sl Proof.} First note that permutation identity is equivalent to the special case 
of the functorial implication when $\rho$ is a base permutation. If the permutation 
identity and (\ref{eq-specpairing}) hold in $T$, then so does the functorial implication 
for injective base morphisms by the proof of Proposition 3.24 on p. 137 in \cite{BEbook}. 
Suppose now that the functorial implication holds for injective base morphisms. 
Then, as noted above, the permutation identity holds.  To prove that 
(\ref{eq-specpairing}) holds, suppose that $f: n \to n + p$ and $g : m \to n + m + p$,
and let $h = \langle f \cdot (\id_n \oplus 0_m \oplus \id_p), g\rangle : n + m \to n + m + p$. 
Then letting $\rho = \id_n \oplus 0_m$, we have 
\begin{eqnarray*}
f\cdot (\rho \oplus \id_p) &=& \rho \cdot h,
\end{eqnarray*} 
so that 
\begin{eqnarray*}
f^\dagger &=& \rho \cdot h^\dagger
\end{eqnarray*} 
by the functorial implication for injective base morphisms. \eop

It is known that the permutation identity holds in a preiteration theory iff 
the block transposition identity does. This is due to the fact that every 
permutation $[n] \to [n]$ can be written as a composition of block transpositions, and 
if the permutation identity holds for bijective base morpisms $\pi_1,\pi_2: n \to n$, 
then it also holds for $\pi_1\cdot \pi_2: n \to n$.

\begin{cor}
\label{cor-Conway1}
A preiteration theory $T$ is a Conway theory iff the 
(base) parameter, fixed point, double dagger and permutation (or block transposition) 
identities and (\ref{eq-specpairing}) hold in $T$. 
\end{cor}

{\sl Proof. } For one direction, recall that one of the axiomatizations of Conway theories consists of the 
left zero, right zero, pairing and permutation identities, cf. p. 212 in \cite{BEbook}. But the left zero identity is 
an instance of the fixed point identity, the right zero identity is an instance of the 
(base) parameter identity and the permutation identity follows from the functorial implication  
for injective base morphisms. Moreover, the functorial implication  
for injective base morphisms follows from the block transposition identity and (\ref{eq-specpairing}) 
as shown in Lemma~\ref{lem-specpairing}. 

For the other direction, recall that all properties mentioned in the
corollary hold in Conway theories. \eop

\begin{cor}
\label{cor-Conway2}
A preiteration theory $T$ is a Conway theory iff the 
(base) parameter, fixed point, double dagger identities 
and the functorial implication for injective base morphisms 
hold in $T$. 
\end{cor}

\begin{cor}
\label{cor-iteration}
A preiteration theory $T$ is an iteration theory iff it satisfies the 
fixed point, (base) parameter and double dagger identities
and the functorial implication for injective base morphisms,
moreover, it satisfies 
\begin{itemize}
\item the commutative identities \cite{Esikaxioms,BEbook} or 
\item the group identities \cite{Esgroup}, or 
\item the generalized power identities \cite{EsAC,EsMSCS}. 
\end{itemize} 
\end{cor}

\begin{cor}
\label{cor-weakiteration}
A preiteration theory $T$ is an iteration theory satisfying the functorial 
implication for base morphisms iff the fixed point, 
(base) parameter and double dagger identities hold in $T$ 
and $T$ satisfies the functorial dagger implication for base morphisms. 
\end{cor}

\end{document}